# Ethical considerations in infectious disease modelling for public health policy: the case of school closures

**Diego S. Silva[1], Sara Y. Del Valle[2], Michael J. Plank[3,4]**

1. Sydney Health Ethics, School of Public Health, University of Sydney, Australia
2. Associate Laboratory Directorate for Global Security, Los Alamos National Laboratory, USA
3. School of Mathematics and Statistics, University of Canterbury, Christchurch, New Zealand.
4. Te Pūnaha Matatini, Centre of Research Excellence in Complex Systems, Auckland, New Zealand.

## Abstract

Mathematical models of infectious diseases are frequently used as a tool to support public health policy and decisions around implementation of interventions such as school closures. However, most publications on policy-relevant modelling lack an ethical framework and do not explicitly consider the ethical implications of the work. This creates a risk that unintended consequences of interventions are overlooked or that models are used to justify decisions that are inconsistent with public health ethics. In this article, we focus on the case study of school closures as a commonly modelled intervention against pandemic influenza, COVID-19, and other infectious disease threats. We briefly review some of the key concepts in public health ethics and describe approaches to modelling the effects of school closures. We then identify a series of ethical considerations involved in modelling school closures. These include accounting for population heterogeneity and inequalities; including a diversity of viewpoints and expertise in model design; considering the distribution of benefits and harms; and model transparency and contextualisation. We conclude with some recommendations to ensure that policy-relevant modelling is consistent with some key ethics values.

1. Introduction

Infectious disease modelling had a prominent influence on government policy in many countries during the COVID-19 pandemic (Hadley et al. 2024). However, relatively few publications on policy-relevant modelling explicitly consider the ethics of this work (Silva, 2023, Smith et al. 2023, Zacherson et al. 2024). Unlike clinical research, there is no regulatory or agreed-upon ethical framework to guide mathematical modelling or its application by policymakers.

Mathematical models are frequently (though not always) presented in technical and ostensibly value-neutral terms (Rennie et al. 2024). Yet public health modelling is rarely, if ever, an entirely objective scientific endeavour. Modelers must make practical and subjective decisions about which processes and variables to include in the model and which sources of evidence to prioritize in informing model assumptions and parameter estimates. The questions that a model seeks to answer are often determined by governments or other decision makers and these are inherently value-laden, thus shaping the range of possible outcomes of a model via values from its outset (Silva, 2023).

If modelling is being done purely as a scientific activity, omission of specific ethical considerations relating to the public health implications of the model (beyond adhering to fundamental research ethics) may be permissible. However, when models are designed specifically to inform public policy decisions, the ethics and value judgements embedded in both the design of the model and presentation of its outputs warrant explicit consideration (Zachreson et al. 2024).

In the context of new pandemic threats, modellers understand that their work provides potential solutions that simplify a complex reality, not just in terms of health impacts, but also in terms of other socioeconomic and political dimensions. In this article, we ask: what responsibilities do modellers, and decision makers who rely on their work, bear when preparing for and responding to infectious disease outbreaks and pandemics?

In §2, we introduce some key concepts in public health ethics to develop a lens through which the limitations of models could be assessed. In §3, we focus on modelling the effects of school closures as an important case study of infectious disease modelling for policy advice. In §4, we detail several ethical considerations involved in modelling school closures. We argue that greater sophistication of models could potentially be achieved through interdisciplinary collaboration with humanities and social science scholars, as well as affected communities. We also discuss the importance of addressing population heterogeneity and inequalities, considering the distribution of benefits and harms in the short and long term, and promoting transparency and contextualisation of model assumptions and results. In §5, we conclude with a series of recommendations for improving the ability of models to support decision-making in a way that is consistent with ethical principles when responding to infectious disease outbreaks.

## 2. Key concepts in public health ethics

Public health can be differentiated from other areas of healthcare, including clinical medicine and medical research, given its emphasis on population level challenges. Moreover, public health is 'public' in that solutions to population-level challenges require collective action (e.g., vaccination campaigns that aim at herd protection or immunity) or collective organization of individual actors (e.g., governments collecting taxes to fund vaccination programs) (Verweij and Dawson 2007).

Infectious disease dynamics can give rise to unique ethical dilemmas. For example, it is possible for a behavioural intervention, such as reducing social contacts, to have a net cost for a single individual, but a net benefit, if adopted by everyone (Gersovitz and Hammer, 2004). This raises questions about whether it is ethical to make such an intervention mandatory (Roberts et al., 2023). Furthermore, so-called trolley problems can arise when deciding whether to implement a public health intervention that is expected to reduce the total number of infections during an epidemic but increase the number in a particular subpopulation (Kollepara et al., 2024).

From an ethics viewpoint, the attention to collective action means reimagining or deemphasizing principles and values usually associated with clinical medicine. For example, the four classical principles of biomedical ethics of Beauchamps and Childress (2013) - i.e., beneficence, nonmaleficence, justice, and autonomy - were created primarily to address challenges at the bedside. If we are to use these principles in the public health context, then they need to be adjusted, or perhaps new principles should be used instead.

Over the last 25 years, the public health ethics literature can be grouped into roughly one of three broad categories. First, some scholars have proposed new ethical principles to guide public health practice, either in general or principles designed for specific purposes. For example, Upshur's (2002) use of the harm principle, least restrictive means, reciprocity, and transparency. Another example is the principles of Dawson et al. (2020) for addressing challenges during COVID-19, which included attention to solidarity, autonomy, and equity. Second, some authors have proposed frameworks or tools to guide decision-making around ethical questions and quandaries, such as those by Childress et al. (2002) and Kass (2004). Third, philosophers have written about the foundations of public health and public health ethics, generally centring on notions of justice and the need to provide greater attention to the needs of vulnerable or marginalised persons and populations (e.g., Daniels (2012), Venkatapuram (2011), Powers and Faden (2008)).

Perhaps the greatest point of departure between public health ethics and other areas of health ethics is the emphasis on the fundamentally relational nature between people, i.e., the collective in-and-of-itself and the collective vis-a-vis individuals. The English-language idiom that 'no person is an island' is central to thinking about public health, whether it be how one person's actions affect others, or how people can coexist, thrive, and be healthy in their community. From this generally agreed-upon starting point, certain key principles or values come to the fore (we acknowledge that the descriptions below are cursory and important details and nuance are absent):

**Justice**: at its most basic and historic origins, justice is simply the idea of giving each person their due, or what they are owed. For example, does society owe greater attention to those in need or those who are most productive? The answer to this and similar questions turn on other equally or more fundamental questions about the nature of society and the state, as well as the nature of need and what people deserve. We can also think about justice in terms of the right distribution of goods and harms (i.e., distributive justice), setting up the right social structures (i.e., social justice), and having fair and legitimate procedures for law and public policy (i.e., procedural justice), to name but a few subcategories.

**Equity**: in the context of health, equity denotes just and acceptable inequalities in access to healthcare or health outcomes, while inequities are unjust inequalities. There are two key ideas to note regarding equity in health: first, not all inequalities are unjust or unfair (e.g., it is trivially true that more Canadians suffer frostbite than Australians per year). Second, claiming that something is equitable or inequitable requires engaging in discussions of justice, or what people are owed. In practice, and as it relates to health, appeals to equity must include descriptions of what factors make an inequality unjust, e.g., Aboriginal persons in Australia have poorer clinical outcomes because of systemic racism and a discounting of local/community knowledge; persons of lower socioeconomic status have less access to nutritious food because public transportation is deprioritized by local governments, etc.

**Solidarity**: the need for collective action requires persons or groups to stand with or beside others for common causes, while aiming to centre the voices of marginalized persons or communities. Sometimes, or often, these acts of solidarity - and solidarity is generally understood to require action - necessitates incurring costs on behalf of others. This can include persons or groups in positions of power standing in solidarity with socio-politically or economically marginalised groups, as well as marginalised persons standing in solidarity together to fight against various forms of corrupt power. In public health, solidarity is used to champion causes of persons or groups with poorer (inequitable) health outcomes due to discrimination or prejudice, e.g., harm reduction efforts, such as safe drug injection sites and needle-exchange programs.

**Freedom and autonomy**: freedom is often understood as having an 'adequate' range of choices, such that one is free from undue restrictions, and ensuring individuals retain agency within the context of enacted public health measures. Critically, freedom is always bound by other people's freedom, i.e., freedom can only be exercised reciprocally with other people's freedom. In other words, other people's right to be free means there exists an intrinsic limit to freedom from the outset. This matters when thinking about public health modelling since instituting various public health measures can mean that different people's freedoms are affected differently and sometimes (though not always or often) in mutually exclusive ways. Autonomy, on the other hand, means to be self-law giving, or in practice, to make one's life as one sees fit. It assumes freedom and choice but is more than that; it is a person's ability to have values and long-term life goals, and to plan according to those values and goals. Autonomy is often understood as highly relational, not only in the sense that one person's autonomy must coexist with that of others (as in the case of freedom), but that our values and goals are shaped in relation to others, including family, friends, and broader social groups. Properly understood, freedom and autonomy are central concepts in public health and public health ethics, since many of the measures we institute to promote public health restrict choices, e.g., quarantine orders or the mandatory use of contact tracing apps; or speed limits and restrictions on the sale of alcohol and tobacco in the context of non-communicable diseases. But critically, public health also promotes autonomy (Wilson, 2021): a life of good health is important as an end-in-itself, perhaps, but also instrumental to attaining many other life goals. Public health measures, like mass vaccination programs, function in the background of our lives but allow us to be more fulsomely autonomous as individuals by protecting population-level health.

In the next section, we briefly review previous approaches to modelling the effect of school closures as a public health intervention during outbreaks. We consider some of the limitations of modelling in respect of the ethics principles described in this section, before moving on to describe some ways these can be addressed in §4.

### 3. Policy case study: school closures

School closures are widely recognized as an important control measure for severe pandemics of airborne viruses, given the congregated nature of schools and their potential to act as transmission hubs (WHO, 2005, Holmberg et al. 2006, Cauchemez et al. 2009, Qualls et al. 2017). Numerous mathematical epidemiological models have explored the use of school closures to mitigate the spread of influenza epidemics and hypothetical influenza pandemics (Halloran et al. 2008). These models have also sought to estimate the economic impacts of such closures (Sadique, 2008, Xue et al. 2012, Dauelsberg et al. 2024, Pangalo et al. 2024) highlighting both their potential effectiveness in reducing transmission and the significant societal disruptions they may cause.

During the COVID-19 pandemic, nearly every country implemented school closures, with the exceptions of Burundi, Tajikistan, and Turkmenistan (Hale et al., 2021). These decisions were predominantly informed by pre-existing research for influenza and real-time modelling studies of school closure impacts (e.g. Ferguson et al., 2020; Lee et al., 2020; Brooks-Pollock et al., 2021). However, while many models effectively demonstrated the benefits of school closures in controlling viral spread, they often lacked ethical frameworks that adequately balanced public health objectives with educational equity and broader societal well-being (Germann et al. 2022). Specifically, models often excluded marginalised populations, inadvertently amplifying educational inequities.

Traditional epidemiological models, such as susceptible-infectious-recovered (SIR) models (Hethcote, 2000), tend to simplify the complexities of real-world interactions by assuming homogeneously mixing populations. This oversimplification often results in overestimating the potential for disease transmission, particularly in heterogeneous settings where individual behaviours and localized contexts play a significant role (Koopman, 2002). By contrast, agent-based models (ABMs) (Tracy et al. 2018) offer a granular representation of transmission dynamics, simulating the individual interactions within specific environments, such as schools, neighbourhoods, and workplaces. These models can account for varying transmission risks based on factors such as age, mobility, occupation, and social networks. These factors make ABMs particularly effective for evaluating targeted interventions, such as partial school closures, staggered schedules, or hybrid learning approaches (Germann et al. 2022). However, ethical considerations such as societal impacts of school closures (e.g., disruptions to education, mental health challenges, parental burden, and exacerbation of inequities) often fall outside their scope. Typically, neither SIR-type models nor ABMs account for the social and economic inequities exacerbated by closures, the harm to children's education and socialization, or for the values and priorities of affected communities. Not only are these challenges likely left implicit within models, but what it means to balance them against the important benefits of school closures (namely, reduced transmission and associated morbidity and mortality) also remains underexplored or merely assumed. As a result, even the most sophisticated models may offer incomplete guidance for policy decisions in complex, ethically charged scenarios. Bridging this gap requires integration of model outputs with qualitative assessments from social scientists, educators, and affected communities.

While mathematical models can provide powerful insights, they must be designed to anticipate unintended consequences, particularly for marginalised groups, and to align with ethical principles. Although students from across the socioeconomic spectrum bore the negative impact of school closures (Mazrekaj et al. 2023), the consequences often disproportionately impacted marginalised groups, such as low-income families, single-parent households, students with disabilities, and those relying on school meals or internet access for learning (Goudeau et al. 2021; AITSL 2021; UNESCO 2022; Galasso and Watts 2022). For instance, school closures can exacerbate educational inequalities, increase mental distress among students and teachers (AITSL 2021, Mazrekaj et al. 2023), disrupt access to critical services (e.g., access to subsidized school breakfasts or lunches), and impose severe economic burdens on families without adequate childcare support (Darmody et al. 2021). Furthermore, the psychological impacts of prolonged closures on students, including social isolation and learning loss, remain difficult to quantify within traditional or agent-based modelling frameworks. For example, McGlynn and Stout (2022) examined the challenges faced by four school districts in the United States, emphasizing the pronounced impact of the digital divide on marginalised communities and its implications for equitable access to education during school closures. Still, these harms need to be balanced against the likely reduction in morbidity and mortality associated with school closures, particularly vulnerable persons often at greater risk to the impacts of the virus, e.g., those who are immunocompromised. In short, models play an important role in trying to better predict how these trade-offs might eventuate.

## 4. Ethical considerations in modelling the effects of school closures

In this section, we describe various ethical considerations as they relate to modelling the effect of school closures as a disease control measure. We highlight issues that need to be considered by modellers and by policymakers using modelling results to inform decisions.

**4.1 Population heterogeneity and inequalities**

To address some of the challenges identified above, particularly those relating to justice and equity, models should where possible incorporate variables relating to socioeconomic deprivation, race, and ethnicity (Ma et al. 2021), household composition (e.g., single-parent or multigenerational households), or occupation type (e.g., essential worker status) (Menkir et al. 2021). These attributes shape individuals' risk exposure, contact networks, ability to comply with interventions (Capasso et al. 2022), and access to resources and healthcare services (Richard and Liptisch, 2024), providing a more realistic and equitable framework for modelling pandemic responses.

In addition, model outputs should expand beyond traditional health metrics, such as cases and deaths, to include indicators such as educational attainment gaps, income and employment losses, and mental health impacts. Furthermore, integrating feedback loops between social and economic stressors and disease dynamics, whilst challenging from a modelling point of view, can help not only to enhance realism but to better align with ethical principles by highlighting inequities and guiding more just interventions.

Finally, models should evaluate alternative interventions, such as improved ventilation, masking, or outdoor learning, to compare their impacts with school closures. This approach can highlight trade-offs between public health outcomes, enabling policymakers to make more informed, balanced decisions. For instance, a model comparing the effectiveness of enhanced ventilation systems versus complete school closures might reveal opportunities to reduce transmission without disproportionately harming marginalised groups. This kind of comparison would also help facilitate solidarity by acknowledging that we need to model better ways of sharing the load of protecting populations.

**4.2 Diversity of viewpoints: community engagement and multidisciplinary model design**

Effective model design requires a multidisciplinary approach that incorporates both technical expertise and diverse community perspectives. Developing mathematical models requires critical decisions about which processes and variables to include and which to exclude, and which health metrics to report (e.g., hospitalisations, deaths, years of lives lost). These decisions are influenced by the objectives of the model, data availability and practical considerations, and involve value judgements. Engaging social scientists, educators, and representatives of affected communities ensures that model assumptions and outputs align with ethical principles, such as equity and autonomy. For example, incorporating parental input into school closure models can uncover challenges and open the discussion for alternative solutions.

In addition to selecting variables and processes, modellers typically also make choices about which scenarios or counterfactuals to include. For example, when modelling school closures, the main scenarios for comparison might be to close schools for some period of time or to keep them open. However, these binary comparisons may overlook more nuanced and actionable policy choices. Additional scenarios could include age-specific closures, staggered reopenings, exceptions for children of essential workers, or keeping schools open while closing hospitality venues. Again, multidisciplinary input alongside close collaboration with policymakers and representation from parents (in this instance) are critical to ensure an appropriate set of scenarios are explored. Without such collaboration, there is a risk that important alternatives may be inadvertently excluded from the analysis. Finally, multidisciplinarity and community engagement might also help promote freedom and autonomy by trying to ensure that the value sets of different peoples are best understood and incorporated so that choices are only curtailed, when necessary, thereby also attending to the relationship between different members of a community.

It is important to consider unintended consequences, which will not be apparent in the model results themselves if the relevant variables or processes are not included. For example, one potential unintended consequence of school closures is that children spend more time with parents or grandparents than they would have if schools remained open. Could this potentially accelerate transmission into more vulnerable groups? Such interactions are often not explicitly represented in models but can have significant implications for both public health and policy outcomes. Other anticipatable harms arising from school closures could include harms related to disruption of education, mental health challenges, and the exacerbation of socioeconomic inequalities.

These types of unintended effects are typically not included in traditional mathematical models due to their complexity and difficulty of quantification. However, their exclusion can lead to an incomplete understanding of the trade-offs associated with school closures (discussed further below). Incorporating these broader effects into mathematical models poses significant challenges, both conceptually and methodologically. For instance, how should educational disruption or mental health harms be quantified, and how should they be weighed against health outcomes like reduced transmission or mortality? While directly integrating such variables into models may not always be feasible, their omission does not absolve decision makers from considering them. Instead, modellers can collaborate with social scientists, psychologists, educators, and parents to ensure these harms are described in detail and presented alongside model outputs. This multidisciplinary community-centred approach ensures that policymakers are fully informed of potential consequences, leaving no room for the justification that 'nobody told us that was going to happen'. It may also help identify ways to design and implement the policy so that the benefits of the intervention can be realised whilst minimising the risk of harm and mitigating anticipatable side-effects.

### 4.3 Transparency and contextualization

Transparency is critical not only for scientific credibility but for ethical accountability. Clearly outlining assumptions, parameter sources, and model limitations enables stakeholders and the public to understand both the strengths and gaps in the analysis. For example, acknowledging excluded variables, such as mental health impacts of school closures, ensures decision makers do not overestimate benefits.

When model results are being communicated to policymakers who are not modelling experts, the responsibilities for transparency and accurate communication of model results are broader than in purely academic research. It is important to ensure that stakeholders have a thorough understanding of model limitations, variables that the model excludes or ignores, and uncertainty in model outputs. This is necessary to avoid non-experts placing undue confidence in a particular model output. If parameter selection or scenario choices need to be refined after looking at preliminary results, modellers should be transparent about how and why this was done.

Effective communication will typically require a plain-language description (written and verbal) of model assumptions and results, and will be supported by a strong and trusted relationship between modelling teams and stakeholder organisations. A range of model descriptions with differing levels of technical detail may be required for different audiences, including technical experts, policy specialists, government decision makers, the media and the general public.

Models frequently present results with some measure of uncertainty such as confidence intervals, credible intervals or sensitivity analyses. Clearly communicating to stakeholders how these should be interpreted can be challenging (McCaw and Plank, 2022), particularly when they relate to a time-dependent quantity being plotted on a graph (Juul et al. 2021). It is also important to be clear about sources of uncertainty that confidence intervals do not capture. Frequently, confidence intervals incorporate uncertainty arising from stochastic variability or uncertainty in model parameters after conditioning on available data. They often do not capture uncertainty arising from potential model misspecification or unforeseen changes that invalidate model assumptions.

Where models are being used to inform controversial and high-stakes decisions, such as school closures, there is an onus on modellers to actively consider ways in which their model outputs could be misinterpreted or taken out of context (inadvertently or deliberately) and take steps to prevent this. This can include an explicit description of conclusions that *cannot* be drawn from model results.

**4.4 Modelling the distribution of benefits and harms: the example of time-scales**

One of the key challenges in responding ethically to any infectious disease outbreak will be balancing the trade-offs of particular interventions, or perhaps more specifically, how to justify the distribution of an intervention's benefits and harms. This is squarely a question of justice. Models can quantify some aspects of these distributions. For example, an age-stratified model can quantify the avoided health burden (measured by metrics such as deaths, years of life lost, or disability-adjusted life years) in each age group. This could be directly compared with an estimate for the age distribution of costs, e.g. quantified by the number of school days missed. Such an approach can help models reflect ethical dimensions of intervention impacts. However, it is worth noting that this process itself requires value judgements to determine which variables to stratify on, and which metrics to use to quantify costs and benefits.

An important case of justice-based considerations for modellers has to do with time-scales of pandemics and outbreaks. The dynamic nature of epidemics means that the choice of time period over which to compare alternative policy options can potentially skew the results. The health benefits from reducing transmission may be immediate but some of the harms from disruption to education may take years to play out. If school closures reduce transmission rates, then the model will typically conclude that this reduces the aggregate health burden in a fixed time period. However, the infections and associated harms may, in reality, be simply delayed beyond the model's time horizon rather than avoided altogether. This could lead the model to overstate the benefits of the intervention or understate the harms, and furthermore, may obfuscate who gets the benefits and who bears the burdens.

There are various ways that modellers can avoid these justice-related challenges associated with time-scales. At a basic level, these include: choosing a longer time horizon or running sensitivity analyses on the choice of time horizon; explicitly modelling what happens after the intervention is removed; reporting the size of the susceptible population as a model output, so that the effect of the intervention on potential future risk is transparent. Ideally, model design will also take account of the strategic aims of the public health response, for example: preventing healthcare system capacity being exceeded; minimising infections until a vaccine becomes available; or eliminating an outbreak so that control measures can subsequently be relaxed. These will affect the important trade-offs, the relevant time frames, and counterfactual scenarios to consider.

## 5. Discussion

An important theme in §4 is the interplay between epidemiological models and ethical principles. Models not only need to be informed by ethics principles and values but also play a critical role in clarifying how to uphold those values in practice. Moreover, they help us understand how subtle system changes can influence the principles that communities and decision makers prioritize. For example, when trying to promote justice, policymakers can use models to predict how school closures during an outbreak affect different parts of a community unequally. As such, the soundness of models becomes imperative not only as a public health tool but as a way of promoting ethical and socially responsive decision-making. While models generally cannot independently determine the right course of action, they play a role in understanding the ethical consequences of alternative policy options. In this context, modellers need ethical input through multidisciplinary collaboration and community engagement to create better, more inclusive models. At the same time, policymakers and communities need robust models to help guide decisions that align with their values (Silva et al 2018).

The lack of explicit ethical frameworks to guide decision-making in a pandemic response is not an issue that is unique to mathematical modelling. For example, New Zealand's Inquiry into COVID-19 (New Zealand Royal Commission, 2024) found that, while decision makers did consider ethical principles, this process was largely done intuitively rather than systematically. The report concluded that there should be more explicit use of ethical principles to "consistently and transparently guide decision makers" (New Zealand Royal Commission, 2024). This recommendation applies to the use of mathematical modelling to support policy-making, where including ethical frameworks can ensure that both models and the broader pandemic response are aligned with societal values.

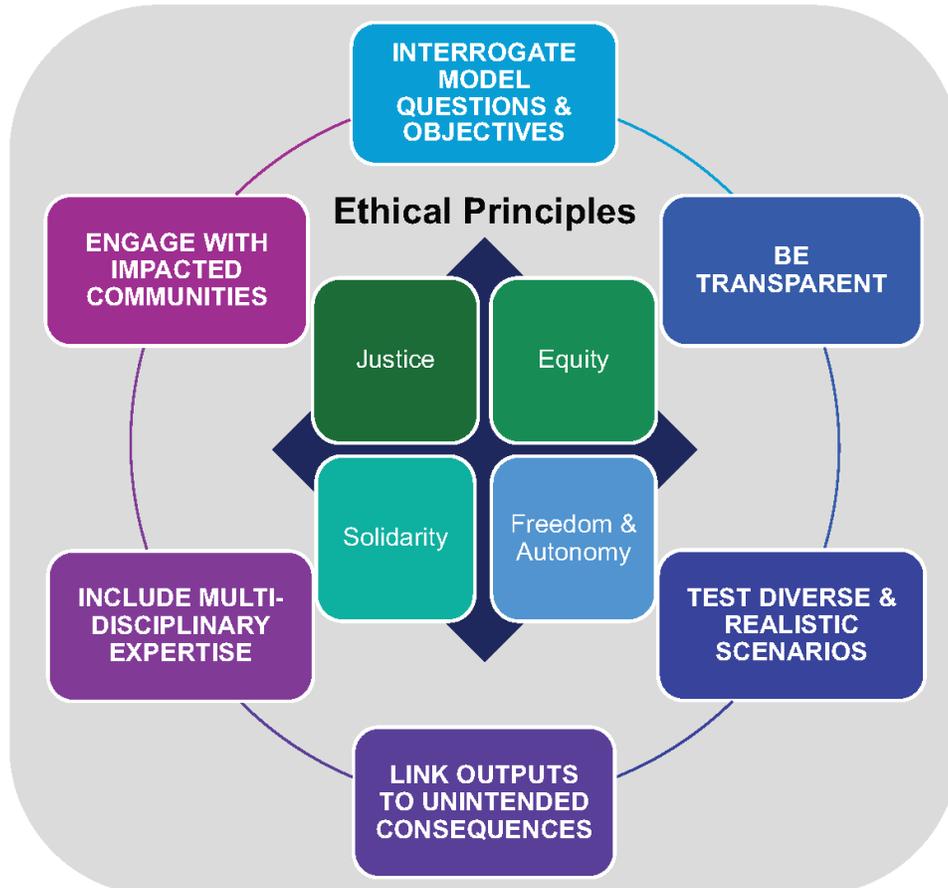

**Figure 1.** Schematic representation of key ethical principles (justice, equity, solidarity, and freedom/autonomy) and the corresponding actions that modelers can take to address these in their models.

Drawing on the ethical issues described in this article, we make some recommendations (see Figure 1 and below) that we believe will improve decisions and ensure that modelling activity is consistent with ethical principles. We do not intend this to be an exhaustive list or a set of "tick boxes", but rather as a series of prompts for modellers to consider in advance of undertaking modelling for policy advice.

- **Include multidisciplinary expertise.** Modellers should partner with experts in disciplines such as social sciences, humanities, and ethics to identify important population characteristics and provide qualitative and quantitative assessments of impacts that cannot be included in the model itself. These better-informed models can then help decision makers better understand the impacts of public health measures on persons freedoms and how to uphold justice in the balancing of various peoples' interests.

- **Engage with impacted communities.** Modelling teams and stakeholders should actively collaborate with representatives from impacted communities. This engagement will help ensure that diverse perspectives and value sets are accounted for, enabling the design of interventions that minimize negative consequences. By fostering these relationships, modellers can better align their models with the needs and priorities of the communities they intend to serve, demonstrating solidarity and a commitment to ethical decision-making.
- **Interrogate model questions and objectives.** The questions that the model seeks to answer may be initially put forward by an external decision maker whose strategic objectives may not always be clear to modelling teams. Modellers should feel able to interrogate and refine model questions in tandem with the decision maker and to seek clarity, as far as is feasible, about their strategic objectives and ensure equity remains at the forefront of decision-making even during emergencies.
- **Be transparent.** Modellers need to be transparent about how the scope and design of a model was chosen and how any post-hoc validation was carried out, clear about the limitations of the model and, where possible, explicit about conclusions that *cannot* be drawn from the model results. Transparency is a key aspect of maintaining and promoting just and fair procedures.
- **Test diverse and realistic scenarios.** A wider range of policy scenarios should be developed and tested to capture the trade-offs between different interventions, ensuring that models reflect realistic and nuanced policy options. Scenarios should also consider the potential short- and long-term impacts of the recommended policies, including their unintended consequences.
- **Link outputs to unintended consequences.** Modellers should focus on developing approaches that explicitly link model outputs to broader societal impacts and the various inequities that different measures might inadvertently introduce. This includes identifying and documenting potential unintended consequences that may arise from the modelled interventions, even if these are not directly quantified within the model.

By adopting these practices, models can provide a more comprehensive basis for policy-making and decision support, providing a nuanced understanding of both the intended outcomes of interventions and their potential risks and trade-offs. The goal is not to make models all-encompassing, but to ensure they are embedded within a broader decision-making process that integrates diverse perspectives and expertise.

Finally, adopting and developing the practices described in this article requires time and a foundation of trust among researchers in different disciplines, as well as between academia, communities, and public health policymakers. Attempting to establish these collaborations during an emergency, when modelling outputs are likely to be required in short time frames to inform urgent decision making, is unlikely to yield optimal outcomes. Therefore, sustained efforts during non-emergency periods are critical to ensure ethical, effective, and well-informed pandemic responses in the future. This requires ongoing investment in building and maintaining modelling capabilities, collecting more detailed social and demographic data to support more nuanced models such as ABMs, and ensuring routine maintenance of both new and existing tools. One concrete step forward is to proactively foster relationships among modellers, social scientists, and public officials. Such interdisciplinary partnerships not only enhance ethical and contextual sophistication of models but can also strengthen advocacy for allocating resources. Justifying these investments in the absence of immediate crisis is difficult but essential if models are to serve as trustworthy and ethically sound tools in future pandemics.

## Acknowledgements

The authors would like to thank Cameron Zachreson for suggesting that we collaborate on this article. The authors are grateful to Joel Miller, Zeb Jamrozik, and one anonymous peer reviewer for feedback on a previous version of this manuscript.

## Funding

MJP was partially supported by Te Niwha Infectious Diseases Research Platform, Institute of Environmental Science and Research, grant number TN/P/24/UoC/MP and the Marsden Fund, grant number 24-UOC-020. SYD was partially funded by the Laboratory Directed Research and Development program of Los Alamos National Laboratory under project number 20240066DR. This work was performed at Los Alamos National Laboratory (LANL), an affirmative action/equal opportunity employer, which is operated by Triad National Security, LLC, for the National Nuclear Security Administration (NNSA) of the United States Department of Energy (DOE) under contract #19FED1916814CKC. This work is approved for public distribution under LA-UR-25-20784. Its contents are solely the responsibility of the authors and do not necessarily represent the official views of the Los Alamos National Laboratory.

## Ethics

Ethics approval was not required for this work.

## Use of Artificial Intelligence (AI) and AI-assisted technologies

We did not use AI or AI-assisted technologies in this work.

## Data, code and materials

No data, code or materials were used in this work.

## Competing interests

The authors declare no competing interest.

**CRediT (Contributor Roles Taxonomy)**

Conceptualization: all
Investigation: all
Visualization: SYD
Writing – original draft: all
Writing – review & editing: all

**References**

1. AITSL – Australian Institute for Teaching and School Leadership. 2021. *The Impact of COVID-19 on Teaching in Australia: A Literature Synthesis*. (accessed May 23, 2025 - aitsl.edu.au/docs/default-source/research-evidence/impact-of-covid-19-literature-synthesis.pdf?sfvrsn=7abdac3c_2).
2. Beauchamp TL, Childress JF. 2013 *Principles of bioethics*. 7th ed. Oxford University Press, Oxford.
3. Brooks-Pollock E, Read JM, McLean AR, Keeling MJ, Danon L. 2021 Mapping social distancing measures to the reproduction number for COVID-19. *Phil. Trans. R. Soc. B* 376, 20200276. (doi:10.1098/rstb.2020.0276)
4. Capasso A, Kim S, Ali SH, Jones AM, DiClemente RJ, Tozan Y. 2022 Employment conditions as barriers to the adoption of COVID-19 mitigation measures: how the COVID-19 pandemic may be deepening health disparities among low-income earners and essential workers in the United States. *BMC Public Health* 22, 870. (doi:0.1186/s12889-022-13259-w)
5. Cauchemez S, Ferguson NM, Wachtel C, Tegnell A, Saour G, Duncan B, Nicoll A. 2009 Closure of schools during an influenza pandemic. *Lancet Infect. Dis.* 9(8), 473-81. (doi:10.1016/S1473-3099(09)70176-8)
6. Childress JF, Faden RR, Gaare RD, Gostin LO, Kahn J, Bonnie RJ, Kass NE, Mastroianni AC, Moreno JD, Nieburg P. 2002 Public health ethics: mapping the terrain. *J. Law Med. Ethics* 30(2), 170-8. (doi:10.1111/j.1748-720x.2002.tb00384.x)
7. Daniels N. 2012 *Just health: meeting health needs fairly*. Cambridge University Press, Cambridge. (doi:10.1017/CBO9780511809514)
8. Dauelsberg LR, Maskery B, Joo H, Germann TC, Del Valle SY, Uzicanin A. 2024 Cost effectiveness of preemptive school closures to mitigate pandemic influenza outbreaks of differing severity in the United States. *BMC Public Health* 24(1), 200. (doi:10.1186/s12889-023-17469-8)
9. Darmody, M., Smyth, E., & Russell, H. (2021). Impacts of the COVID-19 Control Measures on Widening Educational Inequalities. *YOUNG*, *29*(4), 366-380. doi:10.1177/11033088211027412

10. Dawson A, Emanuel EJ, Parker M, Smith MJ, Voo TC. 2020 Key ethical concepts and their application to COVID-19 research. *Public Health Ethics* 13(2), 127-132. (doi:10.1093/phe/phaa017)
11. Fergsuon NM, Laydon D, Nedjati-Gilani G, Imai N, Ainslie K, Baguelin M, Bhatia S, Boonyasiri A, Cucunubá Z, Cuomo-Dannenburg G et al. 2020 Report 9: Impact of non-pharmaceutical interventions (NPIs) to reduce COVID-19 mortality and healthcare demand. Imperial College COVID-19 Response Team Report 9. (doi:10.25561/77482)
12. Galasso I, Watts G. 2022. Inequalities in the Challenges Affecting Children and their Families during COVID-19 with School Closures and Reopenings: A Qualitative Study. *Public Health Ethics* 15(3):240-255.
13. Germann TC, Smith MZ, Dauelsberg LR, Fairchild G, Turton TL, Gorris ME, Ross CW, Ahrens JP, Hemphill DD, Manore CA, Del Valle SY. 2022 Assessing K-12 school reopenings under different COVID-19 spread scenarios – United States, school year 2020/21: a retrospective modeling study. *Epidemics* 41, 100632. (doi:10.1016/j.epidem.2022.100632)
14. Gersovitz M, Hammer JS. 2004 The economic control of infectious diseases. *Econ. J.* 114(492), 1-27. (doi:10.1046/j.0013-0133.2003.0174.x)
15. Goudeau S, Sanrey C, Stanczak A, Manstead A, Darnon C. 2021 Why lockdown and distance learning during the COVID-19 pandemic are likely to increase the social class achievement gap. *Nature Human Behaviour* 5, 1273-1281. (doi:10.1038/s41562-021-01212-7)
16. Hadley L, Rich C, Tasker A, Retif O, Funk S. 2024 How does policy modelling work in practice? A global analysis on the use of epidemiological modelling in health crises. *medRxiv preprint* (doi:10.1101/2024.08.12.24311899)
17. Hale T, Angrist N, Goldszmidt R, Kira B, Petherick A, Phillips T, Webster S, Cameron-Blake E, Hallas L, Majumdar S, Tatlow H. 2021. A global panel database of pandemic policies (Oxford COVID-19 government response tracker). *Nature Human Behaviour* 5(4): 529–538. (doi:10.1038/s41562-021-01079-8).
18. Halloran ME, Ferguson NM, Eubank S, Longini IM, Cumming DAT, Lewis B, Xu S, Fraser C, Vullikanti A, Germann TC et al. 2008 Modeling targeted layered containment of an influenza pandemic in the United States. *Proc. Natl. Acad. Sci.* 105(12), 4639-4644. (doi:10.1073/pnas.0706849105)
19. Hethcote HW. 2000 The mathematics of infectious diseases. *SIAM Rev.* 42(4), 599-653. (doi:10.1137/S0036144500371907)
20. Holmberg SD, Layton CM, Ghneim GS, Wagener DK. 2006 State plans for containment of pandemic influenza. *Emerg. Infect. Dis.* 12(9), 1414-1417. (doi:10.3201/eid1209.060369)
21. Juul JL, Græsbøll K, Christiansen LE, Lehmann S. 2021 Fixed-time descriptive statistics underestimate extremes of epidemic curve ensembles. *Nature Phys.* 17, 5-8. (doi:10.1038/s41567-020-01121-y)
22. Kass NE. 2007 Public health ethics: from foundations and frameworks to justice and global public health. *J. Law Med. Ethics* 32, 232-242. (doi:10.1111/j.1748-720x.2004.tb00470.x)

23. Kollepara PK, Chisholm RH, Kiss IZ, Miller JC. 2024 Ethical dilemma arises from optimizing interventions for epidemics in heterogeneous populations. *J. Roy. Soc. Interface* 21, 20230612. (doi:10.1098/rsif.2023.0612)
24. Koopman J. 2002 Controlling smallpox. *Sci.* 298(5597), 1342-1344. (doi:10.1126/science.1079370 )
25. Lee B, Hanley JP, Nowak S, Bates JHT, Hébert-Dufresne L. 2020 Modeling the impact of school reopening on SARS-CoV-2 transmission using contact structure data from Shanghai. *BMC Public Health* 20, 1713 (doi:10.1186/s12889-020-09799-8)
26. Ma KC, Menkir TF, Kissler S, Grad YH, Lipsitch M. 2021 Modeling the impact of racial and ethnic disparities on COVID-19 epidemic dynamics. *eLife* 10, e66601. (doi:10.7554/eLife.66601)
27. Mazrekaj D, De Witte K. 2023. The Impact of School Closures on Learning and Mental Health of Children: Lessons From the COVID-19 Pandemic. *Perspectives on Psychological Science*. 19(4):686-693.
28. McCaw JM, Plank MJ, 2022. The role of the mathematical sciences in supporting the COVID-19 response in Australia and New Zealand. *ANZIAM J.* 64(4), 315-337. (doi:10.1017/S1446181123000123)
29. McGlynn A, Stout C. Necessity as the Mother of Invention: Attempting to Overcome the Digital Divide During the COVID-19 Pandemic. InThe Color of COVID-19 2022 Jun 16 (pp. 29-45). Routledge. (doi: 10.4324/9781003268710-3)
30. Menkir TF, Jbaily A, Verguet S. 2021 Incorporating equity in infectious disease modeling: case study of a distributional impact framework for measles transmission. *Vaccine* 39(21), 2894–2900. (doi:10.1016/j.vaccine.2021.03.023)
31. New Zealand Royal Commission. 2024 Lessons from COVID-19 to prepare Aotearoa New Zealand for a future pandemic: Main report. https://www.covid19lessons.royalcommission.nz/assets/Report-pdfs/Royal-Commission-COVID-19-Lessons-Learned-MAIN-REPORT-Phase1.pdf
32. Pangallo M, Aleta A, del Rio-Chanona RM, Pichler A, Martín-Corral D, Chinazzi M, Lafond F, Ajelli M, Moro E, Moreno Y, Vespignani A. 2024 The unequal effects of the health–economy trade-off during the COVID-19 pandemic. *Nature Human Behaviour* 8(2), 264-75. (doi:10.1038/s41562-023-01747-x)
33. Powers M, Faden R. 2008 *Social justice: the moral foundations of public health and health policy*. Oxford University Press, Oxford.
34. Qualls N, Levitt A, Kanade N, Wright-Jegede N, Dopson S, Biggerstaff M, Reed C, Uzicanin A. 2017 Community mitigation guidelines to prevent pandemic influenza — United States, 2017. *Morbidity Mortality Weekly Report* 66(RR-1), 1–34. (doi:10.15585/mmwr.rr6601a1)
35. Rennie S, Levintow S, Gilbertson A, Luseno WK. 2024 Ethics of mathematical modeling in public health: the case of medical male circumcision for HIV prevention in Africa. *Public Health Ethics* 17(3), 125-138. (doi:10.1093/phe/phae009)
36. Richard D, Lipsitch M. 2024 What's next: using infectious disease mathematical modelling to address health disparities. *Int. J. Epidem.* 53(1), dyad180. (doi:10.1093/ije/dyad180)

37. Roberts R, Jamrozik E, Heriot GS, Slim AC, Selgelid MJ, Miller JC. 2023 Quantifying the impact of individual and collective compliance with infection control measures for ethical public health policy. *Science Advances* 9(18), eabn7153. (doi:10.1126/sciadv.abn7153)
38. Sadique MZ, Adams EJ, Edmunds WJ. 2008 Estimating the costs of school closure for mitigating an influenza pandemic. *BMC Public Health* 8, 1-7. (doi:10.1186/1471-2458-8-135)
39. Silva DS, Smith MJ, Norman CD. 2018. Systems thinking and ethics in public health: a necessary and mutually beneficial partnership. *Monash Bioeth Rev*. 36(1-4):54-67. doi: 10.1007/s40592-018-0082-1
40. Silva DS. 2023. Reply to Basseal et al.'s "Key lessons from the COVID-19 public health response in Australia". *Lancet Regional Health - Western Pacific* 30, 100629.
41. Smith BT, Warren CM, Rosella LC, et al. Bridging ethics and epidemiology: modelling ethical standards of health equity. *SSM Popul Health* 24:101481 (doi: 10.1016/j.ssmph.2023.101481)
42. Tracy M, Cerdá M, Keyes KM. 2018 Agent-based modeling in public health: current applications and future directions. *Ann. Rev. Public Health* 39(1), 77-94. (doi: 10.1146/annurev-publhealth-040617-014317)
43. UNESCO – Institute for Statistics. 2022. *From Learning Recovery to Education Transformation: Insights and Reflections from the 4th Survey on National Education Responses to COVID-19 School Closures*. Montreal, Canada.
44. Upshur REG. 2002 Principles for the justification of public health intervention. *Can. J. Public Health* 93, 101-103. (doi:10.1007/BF03404547)
45. Vekatapuram S. 2011 *Health justice: an argument from the capabilities approach*. Polity Press, Cambridge.
46. Verweij M, Dawson A. 2007 The meaning of 'public' in 'public health', in *Ethics, Prevention, and Public Health*, Dawson A and Verweij M (eds.) pp. 13-29, Oxford University Press, Oxford.
47. World Health Organization. 2005 *WHO global influenza preparedness plan: the role of WHO and recommendations for national measures before and during pandemics*. No. WHO/CDS/CSR/GIP/2005.5. https://iris.who.int/handle/10665/68998
48. Wilson J. 2021 *Philosophy for public health and public policy: beyond the neglectful state*. Oxford University Press, Oxford.
49. Xue Y, Kristiansen IS, de Blasio BF. 2012 Dynamic modelling of costs and health consequences of school closure during an influenza pandemic. *BMC Public Health* 12, 1-7. (doi: 10.1186/1471-2458-12-962)
50. Zachreson C, Savulescu J, Shearer FM, Plank MJ, Coghlan S, Miller JC, Ainslie KEC, Geard N. 2024 Ethical frameworks should be applied to computational modelling of infectious disease interventions. *PLoS Comp. Biol.* 20(3), e1011933. (doi:10.1371/journal.pcbi.1011933)